\documentclass[submission,copyright,creativecommons]{eptcs}
\providecommand{\event}{LAFM 13} % Name of the event you are submitting to
\usepackage[english]{babel}
\usepackage[latin1]{inputenc}
\usepackage{amsmath}
\usepackage{amsfonts}
\usepackage{amssymb}
\usepackage{enumerate}
\usepackage{stmaryrd}
\usepackage{alltt}
\usepackage{graphicx}
\usepackage[all]{xy}

\usepackage{localdefs} % Local definitions
\usepackage{compos} % Source code for designs

\title{Actions and Events in Concurrent Systems Design}

\author{ %
	Valentin Cassano \qquad Thomas S. E. Maibaum\\%
	\institute{Department of Computing and Software \\ McMaster University \\ Canada \\ \email{cassanv@mcmaster.ca} \email{tom@maibaum.org}}
}

\def\titlerunning{Actions and Events in Concurrent Systems}
\def\authorrunning{V. Cassano \& T. S. E. Maibaum}

\begin{document}

\maketitle

\begin{abstract}
	In this work, having in mind the construction of concurrent systems from components, we discuss the difference between \emph{actions} and \emph{events}. For this discussion, we propose an(other) architecture description language in which actions and events are made explicit in the description of a component and a system. Our work builds from the ideas set forth by the categorical approach to the construction of software based systems from components advocated by Goguen and Burstall, in the context of institutions, and by Fiadeiro and Maibaum, in the context of temporal logic. In this context, we formalize a notion of a component as an element of an \emph{indexed category} and we elicit a notion of a morphism between components as morphisms of this category. Moreover, we elaborate on how this formalization captures, in a convenient manner, the underlying structure of a component and the basic interaction mechanisms for putting components together. Further, we advance some ideas on how certain matters related to the openness and the compositionality of a component/system may be described in terms of classes of morphisms, thus potentially supporting a compositional rely/guarantee reasoning.
\end{abstract}

\section{Introduction}

	That software based systems are complex is, nowadays, an accepted truism of \emph{Software Engineering}. Moreover, because of their different interaction mechanisms, this claim is particularly true of concurrent systems, i.e., systems consisting of several independent parts acting together. In response, among other works, in \cite{Maibaum00} it is argued that decomposition is the only effective method for dealing with the inherent complexity of software based systems. The basic assumption is: since components are more manageable and more easily comprehended, with the aid of suitable structuring mechanisms, the properties of a system can be entailed from the properties of its components.
	
	On this basis, several methodologies for dealing with the modularization and structuring of software based systems from components have been proposed by the software engineering community, of which architectural descriptions are noteworthy (q.v. \cite{Shaw96}). In brief, an architectural description focuses on the overall organization and structure of a system, i.e., it focuses on how a system can be described in terms of its components. In consequence, the emphasis is placed on the distribution of functionalities to the design elements as compared to consideration of data types and algorithms, facilitating the analysis and design of software based systems in the early stages of development.
	
	In this work, we address the organization and structuring of component based concurrent systems from an architectural perspective. Our intention is to progress towards a better understanding of a number of behavioral principles of concurrent systems; in particular, we advance a distinction between actions and events. In that respect, because of its structuring mechanisms, we conceive this work in the same setting as {\CommUnity} (q.v. \cite{Fiadeiro97} and \cite{Fiadeiro05}).
	
	More precisely, {\CommUnity} is an architecture description language in which components and their associated structuring mechanisms have a formal interpretation in categorical terms. In brief, the basic design units, i.e., components, in {\CommUnity} are defined in a guarded command language style, where commands are named by actions, and they become the objects of a given category of systems. Then, the interaction mechanisms between components are based on the identification of actions and variables originating in different components, and they are formulated as morphisms of this given category. Lastly, the description of a system in {\CommUnity} is given as a diagram of this category, indicating how components are meant to interact, and this system is obtained as a co-limit construction of this diagram.
	
	In turn, the interaction mechanisms in {\CommUnity} allow for the behavior of the components involved in the description of a system to synchronize by identifying actions and variables in an architectural description of a system. Thus, e.g., when a system is constructed from its architectural description, those identified actions are merged into corresponding single actions in the resulting system, and similarly for the case of variables. However, we advance that there are actually at least two ways of identifying actions in the description of a system, only one of which is recognized in {\CommUnity}: (i) as being \emph{essentially the same} (i.e., identical), or (ii) as \emph{occurring simultaneously} (i.e., observed in the same state transition). To capture these two ways of identifying actions, we developed a new language (in the style of \CommUnity), called {\Compos}, where commands are named by events observing sets of actions. From an intuitive point of view, while actions denote ``what is a component capable of doing'', events indicate ``occurrences of participating actions''. The intention is to make explicit the idea of actions occurring simultaneously during the same state transition, as opposed to being identical. Then, morphisms between components reflect the following idea: if an event $e_1$ is mapped to an event $e_2$, the set of actions observed by $e_2$ includes the set of actions observed by $e_1$. As a result of making events explicit in the description of a component, we obtain two ways of merging actions from different components in a system description: identifying actions in the description of a system results in these actions being identical from the perspective of the system, whereas identifying the events observing these actions results in these actions being observed as occurring simultaneously at the level of the system. In that sense, concurrency is now supported explicitly within a component, by an event observing a set of actions, and also at the level of the system, by the resulting interaction of components. In this context, we further discuss matters related to the openness and the compositionality of a concurrent system.
	
	This work is organized as follows: In Section \ref{sec:preliminaries}, we introduce some preliminary definitions with the intent to make the reading of this work more self-contained. In Section \ref{sec:component}, we introduce our view of a component in {\Compos}. In Section \ref{sec:morphism}, we turn the discussion to some matters related to openness and compositionality, and we address how systems can be described from this view of a component. In Section \ref{sec:formalization}, we formalize components and their underlying structuring mechanisms in the language of category theory resorting to the notion of an \emph{indexed category} (q.v. \cite{Tarlecki91}). Lastly, we note how an important property of our formalization gives rise to a basic mechanism for building systems from components. In this work, we assume some familiarity with basic concepts of category theory and indexed categories (q.v. \cite{MacLane71} and \cite{Tarlecki91}, respectively). By way of conclusion,  we report on some related work and indicate some further work that we intend to undertake.

\section{Preliminaries}\label{sec:preliminaries}

	In this section, we introduce some basic concepts and results of indexed categories. Although our intent is that of making the reading of this work more self-contained, some definitions and results will be omitted for the sake of brevity; we refer the reader to the complete exposition of indexed categories given in \cite{Tarlecki91}. As mentioned in the introduction, we assume some familiarity with the basic concepts of category theory (q.v. \cite{MacLane71} is the obliged reference source).
	
	In \cite{Tarlecki91}, the concept of an indexed category is presented having observed that families defined over a collection of indices are ubiquitous. Below, we recall the basic definition of an indexed category, and we illustrate this definition with an example (both are borrowed from \cite{Tarlecki91}).
	
	\begin{definition}[Indexed Category]
		Suppose we are given a category $\Ind$; an indexed category over $\Ind$ is a functor $C: \Ind^{op} \rightarrow \Cat$. We call an object $i$ of $\Ind$ an index and $C(i)$ the $i$-th component of $C$. In addition, given a morphism $\sigma: i \rightarrow j$ of $\Ind$, we call $C(\sigma): C(j) \rightarrow C(i)$ the translation functor induced by $\sigma$.
	\end{definition}
	
	Among others, in \cite{Tarlecki91}, the definition of an indexed category is motivated by the following example.
	
	\begin{example}[Many Sorted Sets] Suppose we are given a set $S$, there is a functor category $\SSET(S)$ having as objects functors $X: S \rightarrow \Set$, and as morphisms natural transformations $\eta: X \Rightarrow X'$ under vertical composition. Customarily, we would write $X: S \rightarrow \Set$ as $\langle X_s \rangle_{s \in S}$, where $X_s = X(s)$, (in that sense, $\langle X_s \rangle_{s \in S}$ is a family of sorted sets), and a morphism $\eta: X \Rightarrow X'$ as $\langle \eta_s: X_s \rightarrow X'_s \rangle_{s \in S}$, (i.e., $\langle \eta_s \rangle_{s \in S}$ is a family of functions of sorted sets) .
	
	Moreover, observe that, a mapping $\sigma: S \rightarrow S'$ on sets of indices induces a functor $\SSET(\sigma): \SSET(S') \rightarrow \SSET(S)$ defined as follows: (i) given an object $X': S' \rightarrow \Set$, $\SSET(\sigma)(X') = \sigma; X': S \rightarrow \Set$; and (ii) given a morphism $\eta': X' \Rightarrow X''$, $\SSET(\sigma)(\eta') = \sigma; X' \Rightarrow \sigma; X''$.
	
	On this basis, it is possible to demonstrate that $\SSET: \Set^{op} \rightarrow \Cat$ is an indexed category.
	\end{example}
	
	In turn, an indexed category gives rise to the notion of a ``flattened'' category introduced below.
	
	\begin{definition}[Flattened Category] Suppose we are given an indexed category $C: \Ind^{op} \rightarrow \Cat$, there is a flattened category $\Flat(C)$ defined as follows:
	
	\begin{itemize}
		\item $\langle i, a \rangle$ is an object of $\Flat(C)$ iff $i$ is an object of $\Ind$ and $a$ is an object of $C(i)$.
		\item if $\langle i, a \rangle$ and $\langle j, b \rangle$ are objects in $\Flat(C)$, then, $\langle \sigma, f \rangle: \langle i, a \rangle \rightarrow \langle j, b \rangle$ is a morphism of $\Flat(C)$ iff $\sigma: i \rightarrow j$ is a morphism of $\Ind$ and $f: a \rightarrow C(\sigma)(b)$ is a morphism of $C(i)$.
		\item if $\langle \sigma, f \rangle: \langle i, a \rangle \rightarrow \langle j, b \rangle$ and $\langle \sigma', g \rangle: \langle j, b \rangle \rightarrow \langle k, c \rangle$ are morphisms in $\Flat(C)$, then, the composition of $\langle \sigma, f \rangle$ and $\langle \sigma', g \rangle$ is defined as $\langle \sigma, f \rangle; \langle \sigma', g \rangle = \langle \sigma;\sigma', f;C(\sigma)(g) \rangle: \langle i, a \rangle \rightarrow \langle k, c \rangle$.
	\end{itemize}
	\end{definition}
	
	The following example illustrates an instance of a flattened category.
	
	\begin{example}[Many Sorted Sets] Suppose we are given a functor $\SSET: \Set^{op} \rightarrow \Cat$ defined as above, the flattened category $\SSet = \Flat(\SSET)$ has as objects pairs $\langle S, X: S \rightarrow \Set \rangle$, i.e., sorted sets with an explicitly given sort, and as morphisms pairs $\langle \sigma: S \rightarrow S', \eta: X \Rightarrow \sigma; X' \rangle$, i.e., mappings of sorted sets with an explicit mapping of sorts.
	\end{example}
	
	The conditions under which an indexed category possesses limits and co-limits are also presented in \cite{Tarlecki91}; herein, we omit their exposition for the sake of brevity. The notions of an indexed category and that of a flattened category gain in interest with respect to our intended formalization of components and morphisms of components, respectively. This formalization is introduced in Section \ref{sec:formalization}.

\section{Our view of a Component}\label{sec:component}

	Before entering into a detailed presentation of our framework, let us make explicit our view on what is to be seen as a component of a system. We understand a component as being a modular part of a system which serves as a basic design unit. In this sense, a component ought to possess an encapsulated behavior and a well defined interface. Herein, we take the behavior of a component as being specified in a propositional modal action setting. From our perspective, this view of a component characterizes in a convenient manner certain aspects of a component we deem are worth investigating, i.e., the notion of an action and that of an event. On this basis, in this section, we describe the structure of a component in {\Compos}.

	\subsection{The Structure of a Component}
	
		A component in {\Compos} is comprised of a signature part, denoting the elements we are able to refer to in this component, and a presentation part, describing the behavior of this component.
		
		In this respect, the signature part $\Signature$ of a component is comprised of the following three disjoint sets of names:
		
		\begin{itemize}
			\item A set $P$ comprised of propositional symbols indicating names of variables.
			
			\item  A set $A$ comprised of symbols indicating names of actions.
			
			\item A set $E$ comprised of symbols indicating names of events.
		\end{itemize}
		
		Similarly to what is argued in \cite{Fiadeiro92}, the names of variables indicate information that is state dependent, and the names of actions and events indicate transitions of state. In particular, we assume that names of actions represent what a component is capable of doing, whereas names of events denote state transitions observing sets of actions occurring simultaneously, i.e., synchronization sets. This distinction between actions and events assumes greater interest when discussing interactions between components, a subject that is dealt with in Section \ref{sec:morphism}.
		
		In turn, the presentation part $\Presentation$ of a component, formulated relative to the signature part of this component, is comprised of:
		
		\begin{itemize}
			\item A description of the information that is seen as being state dependent. We assume that this state dependent information is described by sets of sentences (of a propositional language defined over the names of variables of the component under consideration). From our perspective, this assumption reflects an architectural view of a component, where emphasis is placed on functionalities rather than on operational considerations such as data structures.
			
			\item A description of the actions of this component. In this sense, we assume an action is of the form \[ a \: : \: \pi(a) \rightarrow \delta(a) \] where $a$ is the name of an action, $\pi(a)$ is the \emph{prescription} of this action, while $\delta(a)$ is its \emph{description} (q.v. \cite{Khosla87} for the usefulness of this distinction and approach). Intuitively, an action description is to be understood as asserting ``what is known about the consequences of an action occurrence'', whereas the prescription of an action is to be understood as ``under which conditions is an action allowed to occur''.
			
			\item A description of which events observe which actions as occurring simultaneously.
		\end{itemize}
		
		(It may be noted that nothing has been said about the initial values of the variables in $P$. We have chosen to follow \cite{Fiadeiro05} and to take initialization constraints as part of the configuration of a system and not of the component itself.)
		
		Immediately below, we introduce how the previous concepts are structured within a component in {\Compos}.
		
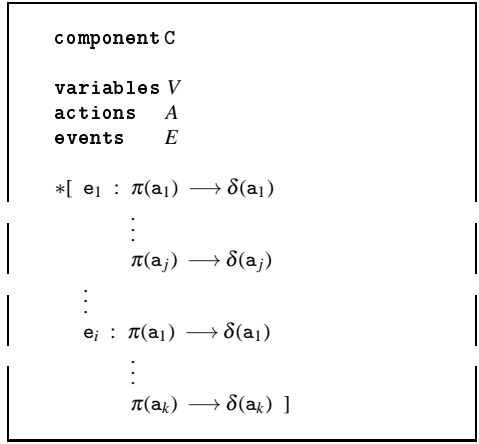
		\begin{component}
		\centering
		\begin{minipage}{6cm}
			\begin{compos}[basicstyle=\scriptsize, frame=tblr]
				component C
				
				variables  \Variables
				actions    \Actions
				events     \Events
				
				*[ e(:$_1$:) : \pi(a(:$_1$:)) --> \delta(a(:$_1$:))
				        (:$\vdots$:)
				        \pi(a(:$_j$:)) --> \delta(a(:$_j$:))
				   (:$\vdots$:)
				   e(:$_i$:) : \pi(a(:$_1$:)) --> \delta(a(:$_1$:))
				        (:$\vdots$:)
				        \pi(a(:$_k$:)) --> \delta(a(:$_k$:)) ]
		\end{compos}
		\end{minipage}
		\end{component}
		
	\subsection{A First Example}\label{sec:example:component}
	
		In this section, we illustrate how an intuitively described scenario can be formalized as a component in {\Compos}. The intention is that of elucidating some of the motivations leading to the definitions presented earlier. For this example, we consider a scenario in which a pilgrim, when hungry, goes hunting for food. In more detail, this activity can be as follows:
		
		\begin{quote}
			\emph{At a given time of the day, a pilgrim becomes hungry and decides to go hunting for food, a turkey in this case. When he spots a turkey, he loads his shotgun, he aims at the turkey, and once the turkey is in his sights, the pilgrim shoots and kills the turkey.}
		\end{quote}

		More precisely, what we wish to illustrate is how this scenario can be described by a collection of actions capturing the behavior of a pilgrim hunting for food. This description is given in \SourceRef{lst:hunting}.
		
		\begin{component}
		\centering
		\begin{minipage}{6cm}
			\begin{compos}[basicstyle=\scriptsize, frame=tblr, title={\captionref[lst:hunting]{Hunting Description}}]
				component Hunting
				
				variables h, l, a, t : bool
				actions bh, ld, am, st, dt
				events e(:$_1$:), e(:$_2$:), e(:$_3$:), e(:$_4$:)
			
				*[ e(:$_1$:): bh: True --> h
				
				   e(:$_2$:): ld: h, \not l --> l
				
				   e(:$_3$:): am: l, \not a --> a
				
				   e(:$_4$:): st: l, a --> \not l, \not a
				       dt: True --> \not t ]
			\end{compos}
		\end{minipage}
% 		\caption{Hunting description} \label{lst:hunting}
		\end{component}
		
		We consider the interpretation of variables, actions and events to be more or less direct. Namely,
		
		\begin{itemize}
			\item We consider the pilgrim being hungry to be denoted by $\mathtt{h}$, the shotgun being loaded and aimed at the turkey to be denoted by $\mathtt{l}$ and $\mathtt{a}$, respectively, and that $\mathtt{t}$ denotes the turkey being alive.
			
			\item For actions, we assume that \lstinline[language=compos]{bh: True --> h} denotes the action of the pilgrim becoming hungry. Intuitively, this action may be understood as an indication that the pilgrim can become hungry at any given time, and that once he becomes hungry, he is indeed hungry. In turn, \lstinline[language=compos]{ld: h, \not l --> l} and \lstinline[language=compos]{am: l, \not a --> a} denote the actions of the pilgrim loading the shotgun and aiming it at the turkey, respectively. The intuitive understanding of these actions is more or less direct. Their prescriptions capture the intention that the pilgrim loads his shotgun when he is hungry, provided the shotgun is unloaded, and that he aims the shotgun at the turkey only if the shotgun is loaded and not already aimed at the turkey he intends to shoot. The intuitive understanding of the description of these actions is immediate. Lastly, \lstinline[language=compos]{st: l, a --> \not l, \not a} and \lstinline[language=compos]{dt: True --> \not t} denote the actions of shooting the shotgun and that of the turkey dying. Once more, their intuitive understanding is immediate.
			
			\item For events, we assume that the action of becoming hungry, and those of loading and aiming the shotgun, respectively, are each observed by a different event. However, we take the shooting of the shotgun and the killing of the turkey to be observed by the same event, i.e., they are conceived as occurring simultaneously (perhaps as an indication of the pilgrim being an accurate shooter).
		\end{itemize}
		
		An initial configuration for this component would contain clauses asserting the pilgrim is not hungry, the shotgun is unloaded and the turkey is alive.
		
\section{Our View of Openness and Compositionality in Concurrent Systems Design}\label{sec:morphism}

	In this work, we subscribe to the view that \emph{openness} and \emph{compositionality} are desired properties of a software based system. Briefly, \emph{openness} amounts to considering a component within an arbitrary environment, while \emph{compositionality} aims at the understanding of a system in terms of its underlying components, understood individually, and the manner in which these components are structured to form systems. There is much active research on these areas and the topics are far from being considered resolved (q.v. \cite{deRoever97}).

	In that respect, we suggest that questions about openness, i.e., \emph{when is a component or a system open?}, should be settled on the basis of signatures. More precisely, insofar as signatures are considered to be denoting the things we can refer to in a component, no assertion can be made about other things affecting the behavior of this component, for if this were the case, they would be referable from the signature of the component. On this basis:
	
	\begin{enumerate}[(i)]
	\item We could view the signature of a component as being comprised of symbols of both the component and an arbitrary environment for this component. Although this view would allow for the elements of the environment of a component to be referred to within this component, we argue that this approach possesses a number of methodological drawbacks. First, considering the signature of the environment of a component to be intertwined with that of a component, amounts to considering an environment as an omnipresent entity. In that respect, an environment for a component is not arbitrary, but it is fixed and underspecified. But more importantly, this approach preempts the understanding of a component locally and as complete in itself, as otherwise imposed by its signature.
	
	\item Alternatively, it is possible to argue that there is an arbitrary environment for a component which cannot be referred to from the signature part of a component. This approach assumes that environments are ethereal entities existing in a different conceptual plane than that of components. In that sense, we claim this approach forces the use of meta level reasoning together with object level reasoning, a feature that is clearly undesirable of any formalism for reasoning about software based systems.
	\end{enumerate}
	
	In any case, these approaches are in contravention of the principle of compositionality. Because of the role played by environments, there is no hope of understanding a system in terms of the local understanding of the components forming this system and the way these components interact. That is, the question then becomes: \emph{when is a formalism compositional?}
	
	Following on from the above discussion, we put forward the idea that the methodologically correct approach for discussing matters related to openness and compositionality is that of manipulating signatures via signature morphisms. More precisely, we observe that signature morphisms serve two important purposes we deem necessary if openness and compositionability are to be considered: (i) they allow the language of a component/system to be extended to describe in what way it is related to an environment, and (ii) they serve as the basic structuring mechanism to build systems from components. Thus, from an intuitive point of view, morphisms of signatures allow for openness to be understood in terms of a class of morphisms indicating how a component is seen as being part of different environments, and for compositionality to be understood along translations of signatures.
	
	Further, we believe that the characterization of openness and compositionality in terms of signature morphisms internalizes a view on rely/guarantee reasoning (q.v. \cite{deRoever97}). In this context, a rely condition may be seen as defining a subclass of environment morphisms. How certain assumptions can be posited over signature morphisms has been discussed in \cite{Dimitrakos98} in the context of logical frameworks; we believe that those ideas may apply here.
	
	\subsection{The Structuring of Concurrent Systems}
	
		It is more or less clear that signatures impose a syntactic restriction on components. Unless we consider the existence of some ethereal entities, an unwelcome situation in software engineering, there are no other things than those we are able to refer to in a signature. Thus, if we intend components to be considered open, it is necessary to extend their signatures to describe in what way they are related to an environment.
		
		In that respect, we have argued that the methodologically correct approach is to extend signatures via signature morphisms. For the notion of a component developed in the previous section, these signature morphisms possess the following structure:
	
% 		\begin{definition}[Signature morphism]\label{def:smorphism}
			\begin{itemize}
% 				\item Suppose we are given two components with signatures $\Signature_1$ and $\Signature_2$, a signature morphism is a tuple $\sigma$ comprised of the following elements:

				\item A mapping $\sigma_{\scriptscriptstyle P}$ of the propositional symbols present in the signature of the source component into the propositional symbols present in the signature of the target component.
				\item A mapping $\sigma_{\scriptscriptstyle A}$ of the action names present in the signature of the source component into the action names present in the signature of the target component.
				\item A mapping $\sigma_{\scriptscriptstyle E}$ of the event names present in the signature of the source component into the event names present in the signature of the target component.
			\end{itemize}
% 		\end{definition}
		
		Immediately, a signature morphism establishes a correspondence between the signatures of the components it involves. The interpretation of this correspondence is more or less direct, i.e., the names of variables of the source component are redesignated as names of variables of the target component, and similarly for the case of action and event names. Moreover, if this signature morphism is not surjective, then, the target component contains names that are not referable from the source component. We interpret these ``additional'' names as the names of variables, actions, or events of a system, or an environment(s).
		
		 Based on this notion of a signature morphism, a structure preserving morphism between components, and a component and its environment(s), is a signature morphism that:
		
		\begin{itemize}
			\item Interprets the actions of the source component into the actions of the target component.
			% More precisely, it interprets the prescription of the actions of the source component into the prescriptions of the actions of the target component, and similarly with the descriptions of actions.
			\item Interprets the observance of events of the source component into the observance of events of the target component.
		\end{itemize}

		Intuitively, we understand a structure preserving morphism between components as describing the extent to which a component is considered to be a part of another, along a translation of signatures. More precisely, a structure preserving morphism ought to indicate how the prescription of an action of the source component corresponds to the prescription of a corresponding action in the target component, once symbols for names have been translated; and similarly for the case of an action description. We assume that these mappings of actions prescriptions and descriptions, respectively, are consistent with the usual notions of superposition (q.v. \cite{Fiadeiro97}). In turn, a structure preserving morphism ought to capture in which way the set of actions observed by an event of the source component is accommodated into a set of actions observed by an event of the target component.
		
		The notion of a structure preserving morphism given above equips us with two ways of manipulating actions when relating components: (i) mappings of actions enable the identification of actions, whereas mappings of events enable the concurrent synchronization of sets of actions. In what follows, we will call a structure preserving morphism a \emph{component morphism}, or a \emph{morphism between components}.
		
		Based on this notion of a morphism between components, we take the description of a system to be given by a collection of components, together with a collection of mappings indicating how these components are intended to interact (given by component morphisms). In this setting, concurrent behavior is supported explicitly within a component and at the level of a system description.
		
		As mentioned, we are also considering that morphisms capture in what way a component ought to interact with an arbitrary environment, i.e., in what way they are considered to be open. From our perspective, this view allows for an internalization of rely/guarantee reasoning, permitting reasoning about environments at the object level in terms of morphisms.
		
		In Section \ref{sec:formalization}, we will show that our proposed characterization of components, morphisms, and the structure of a system in a categorical setting enjoys the necessary properties for building systems from components.
	
	\subsection{A Second Example}
	
		In this section, we illustrate the use of morphisms in the structuring of a system from some given components. Once more, consider the case for the hunting scenario given in Section \ref{sec:example:component}; if this scenario is regarded as a system describing the behavior of a pilgrim going on the hunt for a turkey, then, \SourceRef{lst:hunting} corresponds to a monolithic view of a system. For obvious reasons, monolithic views are not suitable for the design of larger systems. Thus, by way of explanation, we structure the hunting scenario from more basic components. These components describe the behavior of the pilgrim, the shotgun, and the turkey, and are given in \SourceRef{lst:pilgrim}, \SourceRef{lst:shotgun}, and \SourceRef{lst:turkey}, respectively. In turn, the architectural description of the interaction between the components describing the behavior of the hunting scenario is given in Diagram \ref{dia:hunting}.
		
		\begin{component}
		\centering
		\begin{minipage}[t]{5cm}
			\begin{compos}[basicstyle=\scriptsize, frame=tblr, title={\captionref[lst:pilgrim]{Pilgrim}}]
			component Pilgrim
			
			variables h, l, a: bool
			actions bh, ld, am, st
			events e(:$_1$:), e(:$_2$:), e(:$_3$:), e(:$_4$:)
			
			*[ e(:$_1$:): bh: True --> h
			
			   e(:$_2$:): ld: h \and \not l --> l
			
			   e(:$_3$:): am: l \and \not a --> a
			
			   e(:$_4$:): st: l \and a --> \not l \and \not a ]
			\end{compos}
		\end{minipage}
		\hspace{.3cm}
		\begin{minipage}[t]{5cm}
			\begin{compos}[basicstyle=\scriptsize, frame=tblr, title={\captionref[lst:shotgun]{Shotgun}}]
				component Shotgun
				
				variables l: bool
				actions ld
				events e(:$_2$:), e(:$_4$:)
			
				*[ e(:$_2$:): ld: \not l --> l
				
				   e(:$_4$:): st: True --> \not l ]
			\end{compos}
			\begin{compos}[basicstyle=\scriptsize, frame=tblr, title={\captionref[lst:turkey]{Turkey}}]
				component Turkey
				
				variables t: bool
				actions dt
				events e(:$_4$:)
			
				*[ e(:$_4$:): dt: True -> \not t ]
			\end{compos}
		\end{minipage}
		\end{component}
		
		\begin{diagram}[!ht]
		\centerline{
		\framebox{
		{\scriptsize
		\xymatrix@R=7mm@C=6mm{
				&
					\mathtt{G}_1 \ar[dl] \ar[dr] \ar@{.>}[ddr] \ar@{-->}@/_2pc/[dddr]
				&
				&
					\mathtt{G}_2 \ar[dl] \ar[dr] \ar@{.>}[ddl] \ar@{-->}@/^2pc/[dddl]
				&
				\\
					\mathtt{Pilgrim} \ar@{.>}[drr] \ar@{-->}@/_1pc/[ddrr]
				&
				&
					\mathtt{Shotgun} \ar@{.>}[d] \ar@{-->}@/_2pc/[dd]
				&
				&
					\mathtt{Turkey} \ar@{.>}[dll] \ar@{-->}@/^1pc/[ddll]
				\\
				&
				&
					\mathtt{Hunting} \ar@{-->}^{\txt{\large\bf !}}[d]
				&
				&
				\\
				&
					\dots
				&
					\mathtt{Environment}_i
				&
					\dots
				&
				\\
		}}}}
		\caption{System description}\label{dia:hunting}
		\end{diagram}
		
		In order to achieve some simplicity in our exposition, we have assumed that the signature mappings among components correspond to inclusions of signatures. (However, in general, the signatures of the components and that of a system may make use of different symbols.) Then, the Components $\mathtt{G}_1$ and $\mathtt{G}_2$ are trivial components indicating how variables, actions, and event names are to be identified at the level of a system. (These components are seen as \emph{glues} in \cite{Fiadeiro05}.) Intuitively, $\mathtt{G}_1$ synchronizes the behavior of the pilgrim loading the shotgun and that of the shotgun being loaded, formulated in Components \ref{lst:pilgrim} and \ref{lst:shotgun}, respectively, via the inclusion of signatures, i.e., at the level of variables, actions, and events. Similarly, $\mathtt{G}_2$ synchronizes the shotgun being shot and the turkey being killed.
		
		From our perspective, the actions and events describing the loading and the shooting of the shotgun, present in both the pilgrim and the shotgun, are identical actions at the level of the system, i.e., there is no distinction between the pilgrim loading the shotgun and the shotgun being loaded. In contrast, we argue that the actions denoting the shooting and the killing of the turkey are distinct actions at the level of the system; however, by identifying the events observing these actions, we assume that the turkey is killed if they occur simultaneously.
		
		From a Software Engineering point of view, the distinction between identifying actions and events may be justified on the following basis.
		
		\begin{enumerate}[(i)]
		\item When actions of different components are identified in the description of a system, we subscribe to the view that a ``failure'' in executing any of these actions locally in a component causes the same ``error'' to be observed at the level of the system. In the hunting scenario, this amounts to considering that the loading of the shotgun and the shotgun being loaded are essentially the same from the perspective of the system, i.e., it is impossible for the pilgrim to fail in loading the shotgun, and for the shotgun to be ``magically'' loaded.
		
		\item Instead, when events of different components are identified in the description of a system, we subscribe to the view that the sets of actions observed by these events are viewed as occurring simultaneously. In this way, there is scope for a ``failure'' in locally executing an action not to result in the same ``error'' being observed at the level of a system. In the hunting scenario, the previous view amounts to saying that a failure in shooting the shotgun does not necessarily imply that the turkey is still alive, e.g., this turkey may have died from the stress provoked by being shot at.
		\end{enumerate}
		
		In summary, we interpret Diagram \ref{dia:hunting} as follows: (i) solid arrows indicate how components are intended to interact, (ii) dotted arrows denote the system obtained as the result of these components being put together in accordance with their interaction mechanisms (we have shown this system looks like in \SourceRef{lst:hunting}), (iii) dashed arrows describe components as being open, i.e., how they participate in an arbitrary environment.
		
		In addition, since we have in mind that a system should be obtained via a colimit construction, Diagram \ref{dia:hunting} is further interpreted as indicating that, provided that environments respect interactions of components, i.e., provided the outer diagram commutes, the system extends uniquely into this environment. (The latter can be seen as a basic rely condition.)

\section{A Category of Concurrent Systems}\label{sec:formalization}

	In this section, we make precise in which sense components and component morphisms give rise to a category of systems. From our perspective, the conceptual framework provided by category theory conveys a suitable degree of abstractness and simplicity for describing the structuring and the construction of concurrent systems. This view corresponds to the categorical approaches advanced by Goguen and Burstall, in the context of Institutions (q.v. \cite{Goguen91}), and by Fiadeiro and Maibaum, in the context of temporal logic and {\CommUnity} (q.v. \cite{Fiadeiro92} and \cite{Fiadeiro97}, respectively). On this basis, the formalization of components and component morphisms we advance is given in terms of indexed categories (q.v. \cite{Tarlecki91}). In what follows, we elaborate on how this formalization captures our view of a component and that of a component morphism, introduced in Sections \ref{sec:component} and \ref{sec:morphism}, respectively. In particular, the emphasis is placed on eliciting how component morphisms arise naturally from morphisms of signatures. Finally, we note how an important property our formalization of component and component morphisms gives rise to a basic mechanism for building systems from components.
	
	\subsection{The Category Sys}
	
	As mentioned, the categorical formalization we propose is given in terms of indexed categories. In the rest of this section, we explain the motivations underlying the selection of this formalization and in what way it captures components and component morphisms adequately.
	
	Earlier on, in Section \ref{sec:component}, we elaborated on the notion of a component as being comprised of a signature part and a presentation part. Moreover, we commented on how this particular notion of a component is intended to serve as a suitable formal object for studying the difference between identifying actions and events at the level of a system. Below, we provide a categorical formulation of the structure of such a view of a component.
	
	\begin{definition}[Components]\label{def:component}
		With respect to an underlying functor $\Sentences: \Set \rightarrow \Set$, a component is a tuple $\Component = \langle P, A, E, \pi, \delta, \varepsilon \rangle$ where:
		
		\begin{itemize}
			\item $P$, $A$, and $E$ are three disjoint sets of names.
			\item $\pi : A \rightarrow \Set$, $\delta : A \rightarrow \Set$, $\varepsilon : E \rightarrow \Set$ are functors such that the following conditions are satisfied:
		
			\item[--] For all $a \in A$, $\pi(a), \delta(a) \subseteq \Sentences(P)$.
			\item[--] For all $e \in E$, $\epsilon(e) \subseteq A$.
		\end{itemize}
	\end{definition}
	
	In Def \ref{def:component}, the functor $\Sentences$ on sentences provides us with a way of abstracting the structure of a sentence, so that it becomes logic independent. The main idea behind $\Sentences$ is that sets of propositional symbols and mappings between these sets form a category. Then, each set of propositional symbols $P$ is associated with the corresponding set of sentences $\Sentences(P)$ over $P$, in such a way that translations between sets of propositional symbols are mapped to translations of sentences. In particular, $\Sentences$ enables for the manipulation of formal languages under syntactic changes. We find this abstract view of sentences and translations to be useful while manipulating sets of propositional symbols (q.v. \cite{Maibaum97}).
	
	Moreover, for a component $\Component$, the triples $\langle P, A, E \rangle$ and $\langle \pi, \delta, \varepsilon \rangle$ correspond to the notions of a signature and that of a presentation given in Sections \ref{sec:component} and \ref{sec:morphism}, respectively. As usual, $\Set$ is the category of sets and total functions. Thus, because any set defines a discrete category, it follows that, $\pi$, $\delta$ and $\varepsilon$ are functors. In that sense, $\pi$ and $\delta$ denote the prescription and the description of an action, respectively, while $\varepsilon$ denotes the set of actions observed by an event $e$. In turn, the conditions imposed on these functors indicate that the prescription and the description of an action are restricted to what can be referenced by the set of propositional symbols in $P$, and similarly to the previous case, an event can only observe a subset of the actions in $A$ as occurring simultaneously. At this point, it is not difficult to note in which manner signatures impose a syntactical restriction on what can be referenced in a component (cf. Section \ref{sec:morphism}).
	
% 	(cf. Def. \ref{def:presentation}). Finally, it is not difficult to see in what way each functor in $\Presentation$ corresponds to an object of an appropriate flattened (indexed) category. The case of $\varepsilon$ is direct for it fits with the definition of an object of $\SSet$ (q.v. \cite{Tarlecki91}). The definitions describing $\SSet$ can be, \emph{mutatis mutandis}, used to construct a flattened category $\STh$ having functors into theory presentations as objects (instead of functors into sets). Therefore, $\gamma$ and $\delta$ become objects of $\STh$. The morphisms of $\STh$ are defined accordingly.
	
	In summary, Def. \ref{def:component} formalizes, in categorical terms, the notion of a component introduced in Section \ref{sec:component}. However, if we aim to construct a system in terms of components, it becomes necessary to describe in what way those components are intended to interact. In that respect, based on the categorical definition of components given in Def. \ref{def:component}, let us elicit a notion of a morphism of components. Observe that, in order to relate two different components, it is first necessary to indicate how their respective signatures are related. This can be done easily via signature morphisms. It is immediate that signatures of components define a category $\Sig$. This category has triples of sets as objects, and total functions between triples of sets as morphisms, i.e., $\Sig = \Set^{3}$. But two arbitrary components are not necessarily related simply because there is a signature morphism between them; we want to be able to somehow convey a notion of preservation of structure between these components. For instance, for the case of action prescriptions, suppose that $a_1$ and $a_2$ are action names of a source and a target component, respectively, and that $a_1$ is mapped by a signature morphism to $a_2$; intuitively, we would like to view the set of sentences describing the prescription of $a_1$ as being included, modulo translation of its propositional symbols, into the set of sentences describing the prescription of $a_2$. In other words, if $\pi_1(a_1)$ were the prescription of $a_1$ and $\pi_2(a_2)$ that of $a_2$, we would require $\Pow(\Sentences(\sigma))(\pi_1(a_1))$ to be a subset of $\pi_2(a_2)$, where $\sigma$ indicates a translation of propositional symbols. (The latter is essentially the same as considering that the prescription of an action can be strengthened; for observe that if $\bigwedge \Theta$ is the conjunction of all sentences in $\Theta$, then, $\bigwedge \pi_2(a_2) \supset \bigwedge \Pow(\Sentences(\sigma))(\pi_1(a_1))$.) Moreover, we would require a similar condition on the actions observed by events, i.e., if $e_1$ and $e_2$ were event names of a source and a target component, respectively, such that $e_1$ is mapped by a signature morphism to $e_2$, we would require of the set of actions observed by $e_1$ to be included, modulo translation of action names, into the set of actions observed by $e_2$. The previous discussion is made precise in Def. \ref{def:cmorphism}.
	
	\begin{definition}[Component morphisms]\label{def:cmorphism}
		Suppose we are given components $\Component_1 = \langle P_1, A_1, E_1, \pi_1, \delta_1, \varepsilon_1 \rangle$ and $\Component_2 = \langle P_2, A_2, E_2, \pi_2, \delta_2, \varepsilon_2 \rangle$, a component morphism $\tau: \Component_1 \rightarrow \Component_2$ is a tuple $\langle \sigma, \tau_{\pi}, \tau_{\delta}, \tau_{\varepsilon} \rangle$ where:
		
		\begin{itemize}
			\item $\sigma = \langle \sigma_{\scriptscriptstyle P}: P_1 \rightarrow P_2, \sigma_{\scriptscriptstyle A}: A_1 \rightarrow A_2, \sigma_{\scriptscriptstyle E}: E_1 \rightarrow E_2 \rangle$ is a signature morphism.
			\item $\tau_{\pi} = \pi_1 \rightarrow \sigma_{\scriptscriptstyle A};\pi_2$, $\tau_{\delta} = \delta_1 \rightarrow \sigma_{\scriptscriptstyle A};\delta_2$, and $\tau_{\varepsilon} = \varepsilon_1 \rightarrow \sigma_{\scriptscriptstyle E};\varepsilon_2$ are natural transformations satisfying the  following conditions:
			
			\item[--] For all $a \in A_1$, the components of $\tau_{\pi}$ are the domain and the co-domain restriction of $\Sentences(\sigma_{\scriptscriptstyle P})$ to $\pi_1(a)$ and $\pi_2(a)$, respectively; and similarly for the case of $\tau_{\delta}$.
			\item[--] For all $e \in E_1$, the components of $\tau_{\varepsilon}$ are the domain restriction of $\sigma_{\scriptscriptstyle A}$ to $\varepsilon(a)$.
		\end{itemize}
	\end{definition}
	
	Intuitively, observe that the components of the natural transformations of the morphism defined in Def. \ref{def:cmorphism} indicate how the prescriptions and descriptions of the actions, and the actions observed by the events, of a source component are accommodated into a target component. More precisely, $\tau_{\pi}(a): \pi_1(a) \rightarrow \pi_2(\sigma_{\scriptscriptstyle A}(a))$ translates a sentence of $\pi(a)$, given in the language of $P_1$, into a sentence in $\pi_2(\sigma_{\scriptscriptstyle A}(a))$, given in the language of $P_2$, with respect to a translation of propositional symbols $\sigma_{\scriptscriptstyle P}$. Thus, if $\tau_{\pi}(a)$ is not surjective, then, the prescription of an action is strengthened. The situation is similar for $\tau_{\delta}$. In turn, for $e \in E$, $\tau_{\varepsilon}(e): \varepsilon_1(e) \rightarrow \varepsilon_2(\sigma_{\scriptscriptstyle E}(e))$ translates the actions observed in $\varepsilon_1(e)$, given in the language of $A_1$, into the actions observed by $\varepsilon_2(\sigma_{\scriptscriptstyle E}(e))$, given in the language of $A_2$, with respect to $\sigma_{\scriptscriptstyle A}$. If this translation is not surjective, then, the set of actions observed by a source event is extended. In this way, the usual notions of superposition are captured by the property preservation inherent in the definition of a morphism of components.
	
	In a more or less direct manner, the definition of a component and that of a morphism of components given in Defs. \ref{def:component} and \ref{def:cmorphism}, respectively, converge in Proposition \ref{prop:sys}.
	
	\begin{proposition}\label{prop:sys}
		There is a category $\Sys$ having components as objects, and morphisms of components as morphisms (q.v. Defs. \ref{def:component} and \ref{def:cmorphism}, respectively). Moreover, the category $\Sys$ is co-complete.
	\end{proposition}
	
	Proving the first part of Prop. \ref{prop:sys} is somewhat simple once we realize that $\Sys$ corresponds to the flattening of an indexed category (q.v. \cite{Tarlecki91}). In that respect, identity morphisms obviously exist, composition of morphisms is defined pairwise and re-indexing the second component, and to prove that composition is associative is straightforward (q.v. \cite{Tarlecki91} for the details of how this is done). The proof of the co-completeness of $\Sys$ is slightly more complicated and lengthy; however, it follows from the results proven in \cite{Tarlecki91} (in particular, the proof of this part follows an argument similar to that used in the proof of the co-completeness of $\SSet$). In addition, co-limits in $\Sys$ can be computed componentwise. For signatures this is fairly simple; it amounts to computing the co-limits in each one of its constituent categories. For action prescriptions and descriptions, the effect is that of computing the co-limits of their associated sets of sentences along translations of propositional symbols. The process for event names is similar. On this basis, we assume that the architectural description of a system is defined as a diagram into $\Sys$. In turn, we regard that the system is a colimit of this diagram. As elaborated in Section \ref{sec:morphism}, the openness of a component and that of a system defined a class of morphisms, in particular, those that are co-cones of a system description (q.v. Diagram \ref{dia:hunting}).

\section{Our Ideas and Those of CommUnity}

	The formalism we presented is appealing in the context of characterizing and analyzing, from an architectural perspective, concurrent systems built from components where: (i) the behavior of a component, and that of a system, can be described in terms of a collection of actions, and (ii) the interest hinges in the analysis of the interaction mechanisms between actions. There are several architecture description languages fitting into this framework, a survey of which is provided in \cite{Shaw95}. On this basis, we have chosen to develop a formalism that is close in essence to {\CommUnity}.
	
	As mentioned, our interest hinges on {\CommUnity}'s structuring mechanisms. In that respect, {\CommUnity} was originally developed to show how several notions of superposition could be characterized in a categorical setting (q.v. \cite{Fiadeiro97}). Thus, at present (q.v. \cite{Fiadeiro05}), there is only one way of identifying actions in a system description in {\CommUnity}, i.e., through mappings of actions. These mappings of actions are considered in the category of partial functions and in the opposite direction to that presented here. Among others, the rationale for the formalization of mappings of actions in terms of partial functions is motivated by the definition of synchronization sets at the level of the description of a system, i.e., those identified actions in the description of a system are viewed as synchronization sets at the level of a co-limit construction of this system. However, although in a colimit construction the identification of actions is viewed as indicating synchronized sets of actions, at the level of the signature of a component, these synchronization sets are the actions of a component.
	
	We believe that these synchronization sets should be made explicit; thus, in our work events observing sets of actions are included in the signature part of a component. By making events explicit in the signature of a component, we obtain a categorical framework for describing concurrent systems with some worth noticing features, both technically and analytically. From the technical side, dealing with actions and events as in our work, allows us to define morphisms of components in a clear and direct manner as morphisms of an indexed category. From the analytical side, this characterization permits us to study the difference between identifying actions as being equal, and actions occurring simultaneously. We assert that this distinction gains in interest from a software engineering perspective, for it enables a distinction between actions being identified as being the same, and actions occurring simultaneously, i.e., observed in the same state transition (q.v. Section \ref{sec:morphism}).
	
% 	In contrast, in {\CommUnity}, the notion of an event is present at the level of the semantics of actions, but not in the signature of a component (q.v. \cite{Fiadeiro97} and \cite{Fiadeiro05}). In that sense, when compared with {\CommUnity}, our 
	
\section{Conclusions and Further Work}

% 	Normally, are formulated in the language of category theory to study a given problem under consideration.

	Following from the ideas put forward in \cite{Goguen91}, the conceptual apparatus provided by category theory may be seen as a proper foundation for capturing several notions of components and morphisms of components. The same is true of \cite{Fiadeiro92} and \cite{Fiadeiro97}. We consider our work is along these lines of argumentation.
	
	In particular, in this work we have settled the basis for studying what we view are the differences between synchronizing actions and events in the description of a concurrent system. For this purpose, we developed a language in which the distinction between actions and events is made explicit at the level of the syntax of a component. In that respect, the identification of actions from different components in the description of a system can be understood as those identified actions being identical from the perspective of the system (similar to the case of procedure invocation), whereas the identification of events from different components can be understood as an indication of the actions observed by these events occurring simultaneously from the perspective of the system (a kind of conventional parallel composition of actions). From a software engineering point of view, we justify the previous on the following basis: in identifying local actions at the level of a system's description, we ascribe to the view that a failure in executing any of the local actions results in a failure in executing all of these local actions. Instead, in identifying events at the level of a system's description, we subscribe to the view that a failure in executing one of the actions observed in these events does not necessarily entail a failure in all of the actions they observe.
	
	In turn, we presented how the underlying structure of a component can be captured as the objects of an indexed category in a suitable and convenient manner, and we elicited certain interaction mechanisms between components can be captured as morphisms of components, in a way such that a category is formed. This categorical formalization of components and component morphisms also enjoys some of the desired properties for the construction of systems, i.e., co-completeness. We consider this formalization to be more than a mathematical exercise. In particular, since morphisms between components become first class concepts, this categorical formalization enables a study of matters related to the openness and the compositionality of a component or system in terms of classes of morphisms. This discussion was advanced in Section \ref{sec:morphism}. We argue that this last approach simplifies the understanding of a component while maintaining its openness. More precisely, while a component is understood locally, as dictated by its signature, morphisms capture in  what way it is supposed to interact with an arbitrary environment, i.e., in which way it is open. In this regard, we believe that making explicit environments as classes of morphisms internalizes a form of rely/guarantee reasoning. For instance, a rely condition can be viewed as defining a subclass of environment morphisms. Some ideas of how conditions can be imposed on classes of morphisms based on signatures has been discussed in \cite{Dimitrakos98} in the context of extending languages in arbitrary logical settings. We believe that these ideas are worth exploring.
	
	At the same time, we view the ideas herein presented as being a work in progress rather than the final saying, i.e., they deserve further study and development, tasks we intend to undertake. In particular, it remains to complete a theory for reasoning about components and systems. In that respect, it would be interesting to include a notion of violation in the occurrence of an action, so that, in the set of actions observed by an event, we can distinguish failing from non-failing actions. Some ideas of how this can be done in the context of deontic logic are set forth in \cite{Castro09}. Moreover, we intend to give a description of the behavior of components in terms of \textsc{SMV} programs (q.v. \cite{McMillan01}), since this would enable us to perform an automatic analysis of some temporal properties of components. In the context of {\CommUnity}, this is dealt with in \cite{Aguirre07}. Lastly, we deem also necessary to further explore in which way rely/guarantee properties can be formulated in terms of classes of morphisms. In that respect, an interesting result would be to investigate whether different classes of morphisms characterize different rely/guarantee properties.

\bibliography{bibliography}\bibliographystyle{eptcs}

\end{document}